\documentclass[10pt]{article}

\usepackage{amsmath}
\usepackage{amssymb}

\usepackage{graphicx}

\usepackage{cite}

\usepackage{color}

\usepackage[labelfont=bf,labelsep=period,justification=raggedright]{caption}

\makeatletter
\renewcommand{\@biblabel}[1]{\quad#1.}
\makeatother

\date{}

\begin{document}

\begin{flushleft}
{\Large
\textbf{Googling Social Interactions: Web Search Engine Based Social Network Construction}
}
\\
Sang Hoon Lee$^{1,\ast}$,
Pan-Jun Kim$^{2}$,
Yong-Yeol Ahn$^{3,4}$,
Hawoong Jeong$^{1,5,\dagger}$
\\
\bf{1} Department of Physics,
Korea Advanced Institute of Science and Technology, Daejeon 305-701,
Korea
\\
\bf{2} Institute for Genomic Biology,
University of Illinois at Urbana-Champaign,
Urbana, IL 61801, USA
\\
\bf{3} Center for Complex Network Research,
111 Dana Research Center, Northeastern University,
Boston, MA 02115, USA
\\
\bf{4} Center for Cancer Systems Biology (CCSB) and Department of Cancer Biology, Dana-Farber Cancer Institute, Boston, MA 02115, USA
\\
\bf{5} Institute for
the BioCentury, Korea Advanced Institute of Science and Technology, Daejeon 305-701,
Korea
\\
$\ast$ Present address: Department of Physics, Ume{\aa} University, 901 87 Ume{\aa}, Sweden.
E-mail: lshlj82@gmail.com \\
$\dagger$ E-mail: hjeong@kaist.edu
\end{flushleft}

\section*{Abstract}
Social network analysis
has long been an untiring topic of sociology.
However, until the era of information technology, the availability
of data, mainly collected by the traditional method of personal survey,
was highly limited and prevented large-scale analysis.
Recently, the exploding amount of automatically
generated data has completely changed the pattern of research.
For instance, the enormous amount of data from so-called high-throughput
biological experiments has introduced a systematic or network viewpoint
to traditional biology. Then, is ``high-throughput'' sociological
data generation possible? Google, which has become one of the most influential symbols of
the new Internet paradigm within the last ten years, might provide
torrents of data sources for such study in this (now and forthcoming)
digital era.
We investigate social networks between people by extracting
information on the Web and introduce new tools of analysis
of such networks in the context of statistical physics of
complex systems or socio-physics. As a concrete and illustrative example, the members of the 109th
United States Senate are analyzed and it is demonstrated that the methods of construction and
analysis are applicable to various other weighted
networks.

\section*{Introduction}
Social relationships among people~\cite{Wasserman,Freeman1977} are composed of various weight of
ties, as much as metabolic pathways~\cite{Almaas2004}
or airline traffic networks~\cite{Barrat2004b,Barrat2004}.
However, introducing proper weight for the relationships in social networks is
not an easy task
since it is hard to objectively quantify the relatedness among people.
As people's activities on the Web and communications via social networking
service become more popular, information about the social relationships
among people (especially for famous figures, through news and blog sites)
becomes available and can be used as a source of
high-throughput data.
Here, we suggest that the ability of search engines can be used for this task.
Search engines count/estimate the number of webpages including all the
words in a search query, and this feature can be used to measure the
relatedness between pairs of people in social networks in which we are interested.
The more webpages that are found, the more popular or
relevant the combination of the search query is. Therefore, {\em
cooccurrence} of two people in many personal webpages, news
articles, blog articles, Wikipedia, {\em etc.} on the Web
implies that they are more closely related than two random
counterparts.

There are several advantages of using search engines to construct
social relatedness networks. First, with a list of names,
one can systematically count the number of webpages
containing two names simultaneously,
extracted by search engines to assign the weights of all the
possible pairs. This procedure enormously reduces the necessary
efforts to extract social networks, compared with the traditional
methods based on surveys. In addition, such automation makes
analysis of enormous amount of data related to social networks possible
and helps us to avoid subjective bias, such as the ``self-report''
format of personal surveys~\cite{Todd2007}. Furthermore, if one
extracts social networks from a group of people on a regular basis
over a certain period, the temporal change or stability of the
relationship between group members in the period can be monitored.
Although it is possible that some error or artifacts, such as
several people with the same name~\cite{Henzinger2007}, are caused
by this systematic approach, this can also be managed by adding extra
information (such as putting additional queries like the subjects'
occupations into the search engine, in such cases).
Furthermore, the {\em cost} of investigation with the search engine is much smaller.
This example highlights the effectiveness, objectiveness, and accuracy of the usage
of Web search engines.

\section*{Materials and Methods}
\subsection*{Datasets and Google Correlation}
Based on the pairwise correlations extracted from Google, we
constructed and analyzed the weighted social networks among the
Senators in the 109th United States Congress (US Senate), as well as
some other social groups from academics and sports.
Our datasets are three representative communities with very
different characteristics, i.e., politicians, physicists, and
professional baseball players. The US Senate in the 109th
Congress (http://www.senate.gov) consists of $100$ Senators, two for
each state. Among the physicists who submitted abstracts to American
Physical Society (APS) March Meeting 2006~\cite{APSSource}, we
selected the subset of $1143$ authors who submitted more than two
abstracts for computational tractability. Finally, the list of Major League
Baseball (MLB) players is the 40-man roster (March 28, 2006) with
$1175$ players (http://mlb.com). To avoid the ambiguous situation
where there is more than one person with the same name,
the following distinguishing words or phrases were added to
all the search queries for each group: the words are ``senator'' for US Senators,
``physicist'' for APS authors, and ``baseball'' for MLB players.
First, we recorded the number of pages searched using Google for each member's
name, which were assigned as the Google hits~\cite{Simkin2003} showing the fame
of each individual member.

The {\em Google correlation} between two members of a group is defined as
the number of pages searched using Google when the pair of
members' names (and the additional word) is
entered as the search query~\cite{Filtering}.
In this case,
Google shows the number of searched pages including {\em all the
words in the search query}. Simply, this Google correlation value is
assigned as the link's weight for the pair of nodes. If no searched
page is found for a pair, the pair is not considered to be
connected.
Note that the idea of using co-occurrence to quantify the correlation
was presented before in systems biology~\cite{Cohen2005} or
linguistics~\cite{cooccurrence_website,Ferret2004}, but
our work comprehensively approaches such a general concept and
focuses on the digital records to extract information.
The constructed weighted networks are usually densely
connected: the link density, defined as the ratio of existing links
to all the possible links among nodes ($N(N-1)/2$, where $N$ is the
number of nodes), is $0.95$ for the US Senate, $0.16$ for APS
authors, and $0.66$ for MLB players.

\begin{figure}[!ht]
\begin{center}
\includegraphics[width=0.9\textwidth]{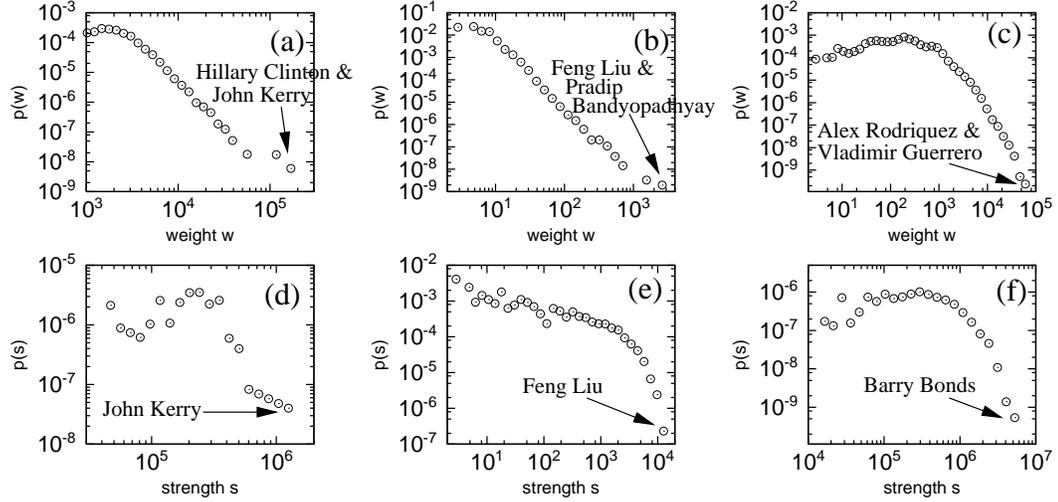}
\end{center}
\caption{
{\bf The weight and strength distributions.}
Google correlation value (weight) distributions $p(w)$
for (a) US Senate, (b) APS authors, and (c) MLB players
and the strength distributions $p(s)$ for
(d) US Senate, (e) APS authors, and (f) MLB players are shown.
Pairs with the largest Google correlation values (a)-(c) and the nodes with
largest strengths (d)-(f) for each plot are indicated.
}
\label{WeightStrength}
\end{figure}

Due to the high link density, elaborating on the weights of links or the
strength (the sum of the weights around a specific node)
of nodes to extract useful information is more
important. Figure~\ref{WeightStrength} shows the weight and
strength distributions for the weighted networks constructed by
assigning the Google correlation values as link weights.
Previous studies on other weighted networks show heavy tailed weight
and strength distributions~\cite{Barrat2004b,Barrat2004}
and our networks also reveal such broad distributions spanning
several orders of magnitude, although the details are different for
each network.

\subsection*{The R{\'e}nyi Disparity}

\begin{figure}[!ht]
\begin{center}
\includegraphics[width=0.5\textwidth]{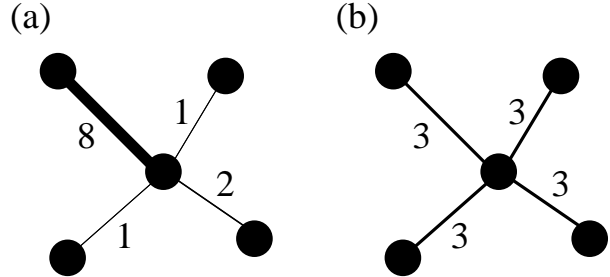}
\end{center}
\caption{
{\bf Two nodes in weighted networks with the same values of
degree and strength.} The degree of the central node in both (a) and
(b) is 4 and the strength is 12, but the distributions of weights
around the nodes are quite different.
}
\label{disparity_example}
\end{figure}

The degree and strength are basic quantities that estimate the
importance of nodes in a weighted
network~\cite{Barrat2004b,Barrat2004}. However, the weights on the links
of two nodes with the same degree and strength
are not necessarily identically distributed. In other words,
just the number of links a node has (degree) and the sum of weights
on the links the node has (strength) are not sufficient to fully
conceive the node's character. For example, two central nodes in
Fig.~\ref{disparity_example} have the exactly same values of degree
and strength, but the weight distributions around the nodes are
totally different. Quantifying such different forms of weight
distributions is important because it can distinguish whether a
node's relationship with its neighboring nodes is dominated only by a small
portion of neighbors
or if almost all the neighbors
contribute similarly to the node's relationship. As an initial step to further
investigation we are interested in the {\em dispersion} or {\em
heterogeneity} of weights a node bears. Although this concept of
disparity is not a new
one~\cite{Almaas2004,Derrida1987,Barthelemy2005}, we suggest a more
general framework of such quantities based on information theory.

Suppose a node $i$ has $k_i$ links whose weights are given by the
set $\{ w_{ij} ~|~ j \in \nu_i \}$, where $\nu_i$ is the set of the
node $i$'s neighboring nodes. The strength of the node is defined as
$s_i = \sum_{j \in \nu_i } w_{ij}$. Now, let us denote
$\tilde{w}_{ij} = w_{ij} / s_i$ for each weight $w_{ij}$ as the
normalized weight.
In the continuum limit of neighbor indices
$x$ sorted by descending weights (without loss of generality)
around the node $i$ whose set of weights is $\{ \tilde{w} (x)
\}$, (the normalization condition becomes simply $\int dx ~ \tilde{w} (x) =
1$ in this case) if all the neighbor indices are re-scaled as $x \to
x' = cx$ (meaning the entire network gets larger by the factor of $c$,
and the normalized weights become $\tilde{w}'(x') = \tilde{w}(x)/c = \tilde{w}(x' / c)/c$
due to the normalization condition),
the quantity $D [ \{ \tilde{w} (x) \} ]$ characterizing the
dispersion of weights should be scaled as $D[ \{ \tilde{w}'(x')
\} ] = c D[ \{ \tilde{w} (x) \} ]$. This scaling behavior is the
same as the degree measure and, in fact, if all the weights are
identical, the quantity is set to precisely become the degree.
We have found a class of
solutions satisfying such scaling conditions, which is the
weighted sum
\begin{equation}
D_i (\alpha) = \left( \sum_{j \in \nu_i} \tilde{w}_{ij}^{\alpha}
\right)^{1/(1-\alpha)} \label{WeightedSum}
\end{equation}
to node $i$, where the constant $\alpha$ is a tunable parameter, and
we denote this measure as the {\em R{\'e}nyi disparity}. If all the
weights are equal, $D_i (\alpha) = k_i$, which is just the degree of
node $i$, regardless of the value $\alpha$. As the weight
distribution deviates from the uniform distribution, $D_i (\alpha)$
also deviates from the degree, the details of which depend on the
parameter $\alpha$, of course. We will use this weighted sum $D_i
(\alpha)$ as the measure of the heterogeneity in the weight
distribution for each node. Note that the logarithm of
Eq.~(\ref{WeightedSum}), $\log D_i (\alpha)$, coincides with the
R{\'e}nyi entropy~\cite{Zyczkowski2003} in information theory, from
which the name ``R{\'e}nyi disparity'' originates.
We have yet to decide the parameter $\alpha$ for $D_i (\alpha)$. In
previous works~\cite{Almaas2004,Derrida1987,Barthelemy2005}, the
quantity called disparity $Y_i$ was defined for each node $i$. Its
scaling behavior is that $Y_i \sim 1/k_{i}$ if the weights are
uniformly distributed and $Y_i \sim \textrm{constant}$ if the weight
distribution is severely heterogeneous. It is easy to see that the
disparity $Y_i$ in
Refs.~\cite{Almaas2004,Derrida1987,Barthelemy2005} is the reciprocal
of a special case of our R{\'e}nyi disparity, with the parameter
$\alpha = 2$, i.e.,
\begin{equation}
Y_i = \frac{1}{D_i (\alpha = 2)} = \sum_{j \in \nu_i}
\tilde{w}_{ij}^2. \label{Y_measure}
\end{equation}
The logarithm of this $D_i (2)$ is also a special case of R{\'e}nyi
entropy, called the extension
entropy~\cite{Zyczkowski2003,Pipek1992} and $Y_i$ is related to the
simple variance $\textrm{Var}(\tilde{w}_{ij})$ by
$\textrm{Var}(\tilde{w}_{ij}) = (Y_i - 1)/k_i$.

If we consider the limiting case of $\alpha \to 1$, we denote it as the
Shannon disparity $D_{\textrm{Shannon}}^{(i)} = \lim_{\alpha \to 1}
D_i (\alpha)$ of the node $i$. In this limit, one can easily verify
that
\begin{equation}
D_{\textrm{Shannon}}^{(i)} = \exp(-\sum_{j \in \nu_i} \tilde{w}_{ij}
\log \tilde{w}_{ij} ) = \prod_{j \in \nu_i}
\tilde{w}_{ij}^{-\tilde{w}_{ij}}. \label{ShannonDisparity}
\end{equation}
One can immediately notice that the Shannon disparity is the
exponential of an even more familiar and widely accepted entropy in
information theory, which is the Shannon
entropy~\cite{Zyczkowski2003}. The scaling property of
$D_{\textrm{Shannon}}$ is similar to $1 / Y$ in
Eq.~(\ref{Y_measure}) and, in fact, for our three weighted networks
the two quantities $D_{\textrm{Shannon}}$ and $1 / Y$ are highly
correlated: the Pearson correlation coefficients are $0.95$ for US
Senate, $0.97$ for APS authors, and $0.96$ for MLB players.

Even though $D_{\textrm{Shannon}}$
and $1 / Y$ are highly correlated in our example networks,
Shannon disparity works better for inhomogeneous weight
distribution than the R{\'e}nyi disparity with $\alpha \ne 1$.
Suppose the weight around a node follows the power-law relation
$\tilde{w}(x) = (\gamma - 1) x^{-\gamma}$ for $x
> 1$, where $x$ is the continuous version of the neighbor indices
sorted by descending weights and the constant $(\gamma - 1)$ is set
to the normalization condition $\int_1^{\infty} dx ~ \tilde{w}(x) =
1$.

\begin{figure}[!ht]
\begin{center}
\includegraphics[width=0.5\textwidth]{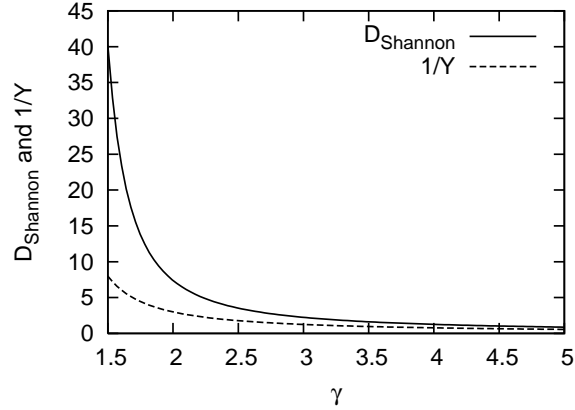}
\end{center}
\caption{{\bf The functional form of $D_{\textrm{Shannon}} = \lim_{\alpha
\to 1} D (\alpha)$ and $1 / Y = D (\alpha = 2)$ in case of the
power-law weight-index relation $\tilde{w}(x) \sim x^{-\gamma}$,
from Eqs.~(\ref{Shannon_explicit}) and (\ref{D_alpha_explicit}).}}
\label{Renyi_power}
\end{figure}

In this continuum limit, we can explicitly
calculate the dependence of $D(\alpha)$ on the power-law exponent
$\gamma$ by direct integration, which is $D(\alpha) =
[\int_{1}^{\infty} dx (\gamma - 1) x^{-\alpha \gamma}
]^{1/(1-\alpha)}$.
The integration is straightforward and the result is
\begin{equation}
D_{\textrm{Shannon}} = \lim_{\alpha \to 1} D (\alpha) =
\frac{1}{\gamma-1} \exp \left( \frac{\gamma}{\gamma - 1} \right)
\label{Shannon_explicit}
\end{equation}
\begin{equation}
D(\alpha > 1) = \left[ \frac{(\gamma - 1)^{\alpha}}{\alpha \gamma -
1} \right]^{1/(1-\alpha)} .
\label{D_alpha_explicit}
\end{equation}
As shown in Fig.~\ref{Renyi_power}, the Shannon disparity
$D_{\textrm{Shannon}}$ is the only R{\'e}nyi disparity showing the
non-polynomial scaling and more sensitive to the exponent $\gamma$
than $D (\alpha > 1)$,
especially when $\gamma$ becomes smaller and $D_{\textrm{Shannon}}$
diverges much faster as $\gamma \to 1$ (the most homogeneous weight
distribution).

\begin{figure}[!ht]
\begin{center}
\includegraphics[width=0.5\textwidth]{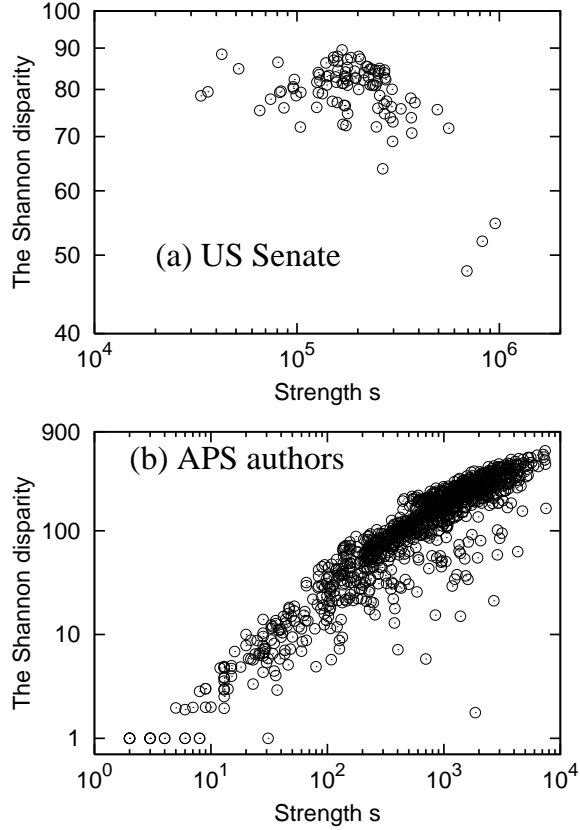}
\end{center}
\caption{
{\bf The scattered plots for the correlation between the
strength $s$ and the Shannon disparity $D_{\textrm{Shannon}}$ of
each node.} The correlations for (a) US Senate and (b) APS authors Google correlation
network are shown. Graphs are drawn in the double logarithmic scale for easy
visualization.
}
\label{StrShannon}
\end{figure}

Figure~\ref{StrShannon} shows the correlation between the strength
$s$ and the Shannon disparity $D_{\textrm{Shannon}}$ of each node
for the two representative cases of the US Senate and APS authors.
From the result, we can conclude that there are some senators with
the very large strength
and very heterogeneous Google
correlation values with other senators, whereas the strength and the
Shannon disparity is positively correlated for APS authors, which
reflects the different attributes of political and academic
communities. For instance, even the politicians meeting many other politicians
may need to focus on the relationships with small groups of others
sharing common interests with them, which might be related to
the partisan politics~\cite{Porter2006,Porter2006a}.

\subsection*{Maximum Relatedness Subnetwork}
As previously stated, the link density values of the Google
correlation networks can be quite large compared with many
sparse networks that have been previously investigated. Especially for the US
Senate network, where almost every member is famous enough to
appear on numerous webpages, almost all the possible pairs of
senators are connected (a single webpage that is searched for each pair
of senators can establish the link between any two senators). In
such a case, beside the statistical properties, such as weight and
strength distributions presented earlier, the mere figure of the
weighted network itself can hardly give any visual clue for specific
information about the structure of the community. In other words,
there exist non-zero correlation values for almost all the pairs of
nodes.

Econophysics has encountered similar situations in dealing with the
financial time series correlations between companies or countries
quite often and one way to circumvent the problem is the famous
maximum (or minimum, depending on the definition of the correlation)
spanning tree (MST)~\cite{MST}. MST extracts the connected subtree
(subnetwork without any loop) which maximizes (or minimizes) the sum
of the weights on all the extracted links and one of the most
popular methods of analyzing time series correlations in
econophysics. Even for an unweighted network, one can extract MST of
the network by assigning the edge betweenness centrality values as
the links' weights so that the ``skeleton'' of the network is
constructed~\cite{DHKim2004a}.

In spite of the popularity of MST and its ability to select
important interactions in many systems composed of pairwise
correlations, there are a few drawbacks in the MST approach. First,
the essential interactions do not need to connect all nodes as one
giant component. In addition, MST uses the global rank of the
weights as prime information for construction, and this might not be
appropriate to access locally important interactions from the individual
nodes' perspective.

We suggest a new approach, called the maximum relatedness subnetwork
(MRS), as an alternative method to extract the essential interactions,
instead of the conventional approach based on maximum spanning tree (MST)~\cite{MST}.
In MRS, for each node $i$, a directed link is connected from the
node $i$ to the other node $j$ with which the node $i$ has its
maximum correlation value. It is possible for a node to have more
than one directed link in the case of the multiple nodes with the same
maximum correlation value. In this way, for a network with an
exactly uniform weight distribution, MRS is restored to the original
network. MRS can resolve the problems of MST by not posing
the restriction of ``one connected component'' and by using the
locally maximum correlation values. Although it is difficult to
assign intuitive meaning to MST, MRS has the clear interpretation of
consecutively connecting to the maximally related nodes. For
instance, a node's incoming degree in MRS shows how many of its
neighbors consider the node as their most important partner and can
be used as the measure of {\em reputation} or importance in the
entire system. Furthermore, the directionality of MRS can yield new
information about the asymmetry of the node pairs which is described
below in detail.

The weighted social networks of our datasets constructed by the
Google correlation values consist of undirected edges, as do
most other social networks in the literature. This bidirectionality
represents the mutual relationship in social networks and is easily
understandable. The ``mutual'' relation, however, may not hold for
the relationship given by the Google correlation. For example, the
fact that a very famous person is connected with many members does
not necessarily mean that she has many friends. Instead, it is
possible that the members became connected to her just because she is
famous and appears on many different webpages. Therefore, many
asymmetric relationships (A is famous {\em mainly} because of B, but
the opposite is not necessarily true) might appear, in the
similar sense of the dependence relation between two authors in the
collaboration network discussed in Ref.~\cite{TZhou2007a}. We
believe that the directionality of MRS represents such asymmetric
relationships or structures. For instance, if we consecutively
``follow'' the directed links in MRS, we can hierarchically reach
links in the ascending order of weights. The link corresponding to
the largest weight should be bidirectional by definition, although
the converse is not always true. In addition, one can extend this
concept further so that each node selects different number of nodes.
One idea is that considering the R{\'e}nyi
disparity from the previous section as ``effective'' degree $D_\textrm{eff}$
and choosing $D_\textrm{eff}$ number of links with largest normalized weights.

\section*{Results}
\subsection*{Maximum Relatedness Subnetwork of US Senate Network}

\begin{figure}[!ht]
\begin{center}
\includegraphics[width=0.9\textwidth]{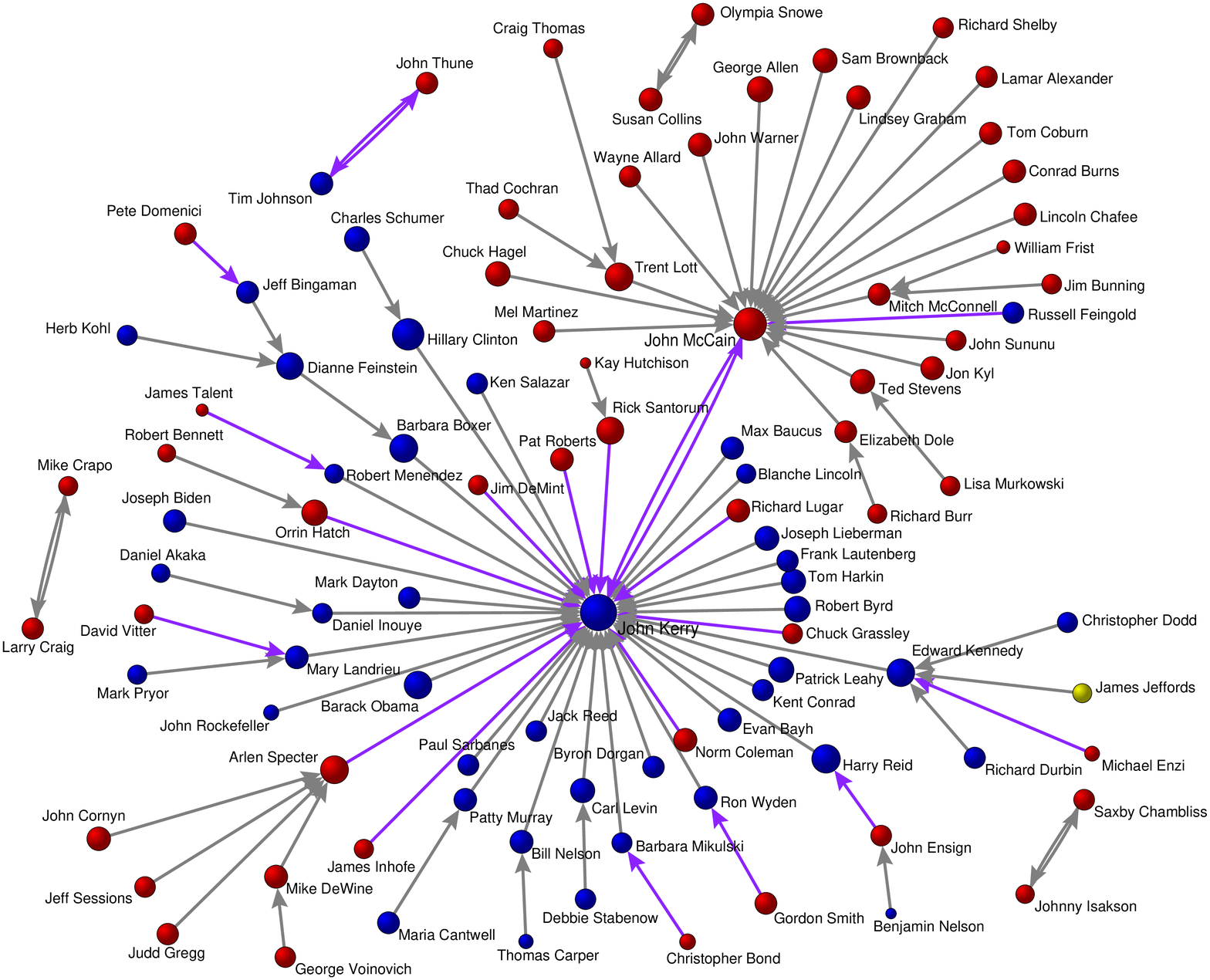}
\end{center}
\caption{
{\bf MRS of the US Senate Google correlation network, with the
Google correlation values for May 4, 2006.} The size of each node is
proportional to the logarithm of the Google hit
value~\cite{Simkin2003}. The nodes' colors represent the political
parties, i.e., blue for the Democratic party, red for the Republican
party, and yellow for the independent Senator James Jeffords.
The links are distinctly colored as positive (gray
links) and negative (purple links) vote correlation
in Eq.~(\ref{VoteCorrelation})
}
\label{Senator_MRS}
\end{figure}

Figure~\ref{Senator_MRS} shows the MRS of US Senate in the 109th Congress.
The most prominent
senators are John Kerry and John McCain, who get many incoming links
from other senators, implying that those numerous senators have the
maximum Google correlation value with Senator Kerry or McCain. The
division or community structure, reasonably consistent with the
senators' political parties, is observed around the two prominent
senators. Another property of MRS is that two adjacent senators are
likely of the same state, e.g., Hillary Clinton and Charles Schumer
from New York, George Voinovich and Mike DeWine from Ohio,
{\em etc}~\cite{SameState}.
All the four ``isolated''
mutually connected pairs are of this case: Johnny Isakson and Saxby
Chambliss from Georgia, Mike Crapo and Larry Craig from Idaho, Susan
Collins and Olympia Snowe from Maine, and Tim Johnson and John Thune
from South Dakota. The last case, Tim Johnson and John Thune from
South Dakota, is especially interesting because those two senators
are mutually connected despite their difference in political
parties. Therefore, MRS is not a random subset of a fully connected
network but represents actual/relevant relationship between people.
Some previous works about the community structures and interpretations
for social networks among politicians
are discussed in Ref.~\cite{Porter2006,Porter2006a}. We also
successfully capture some aspects of this political network,
and present from now on.

One can readily notice that almost all the senators around John
McCain are the Republicans~\cite{Feingold}, whereas
a relatively considerable number of non-Democratic senators are in
John Kerry's side. The likely connection between senators of the
same state can explain such different compositions of communities.
Among the 50 states, 21 states have two Republican senators, 15
states have two Democratic senators, and 13 states have one
Republican and one Democratic senator.
Therefore, a Democratic senator more likely serves with a Republican
senator in a state than vice-versa, which can cause this
kind of community structure. We consider the main factors setting
the structure of MRS as the combination of the ``global'' effect
based on the political parties and senators' individual fame, and
the ``local'' effect based on the home states.

In this paper so far, we have focused on a snapshot of the Google correlation network.
However, we can easily monitor the temporal changes by
constructing the network on a regular basis, which is actually one
of the most important advantages of our network construction scheme. In the
following section, we use the US Senate network once again as an
example of observing structural changes over time near an enormous
political event, the United States Senate elections of 2006.

\subsection* {Temporal Change of the US Senate Network near Election 2006}

\begin{figure}[!ht]
\begin{center}
\includegraphics[width=0.9\textwidth]{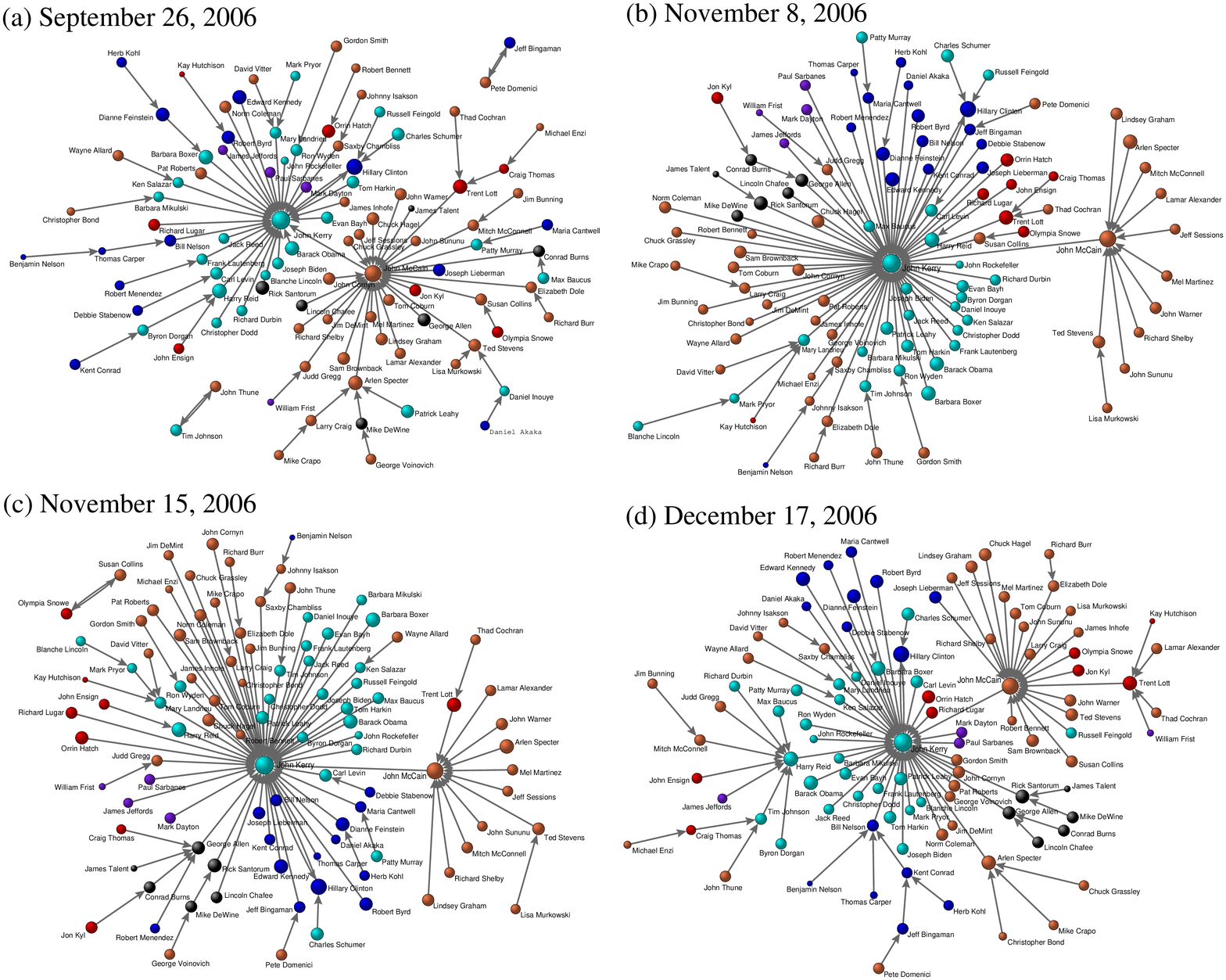}
\end{center}
\caption{
{\bf Four snapshots of MRS of the US Senate Google correlation
network, near United States Senate elections 2006.} The size of each
node is proportional to the logarithm of the Google hit
value. Senators are classified as re-elected
Democrats (dark blue), Democrats not participating in the election
(light blue), re-elected Republicans (dark red), Republicans not
participating in the election (light red), Senators who failed to be
re-elected (black; all Republicans), and Senators who retired
(purple).
}
\label{ElectionUnited}
\end{figure}

The United States Senate elections were held on November 7, 2006. We
expected significant structural changes during this enormous political
event, so we took four snapshots (September 26, November 8, November
15, and December 17) of the US Senate Google correlation networks near
the elections, collecting the Google correlation values
on the four specific days.
Again, we observed the MRS of the network to infer the
structural modification since the overall statistical properties
such as weight and strength distributions, are similar for the four
data. In Fig.~\ref{ElectionUnited}, we present four snapshots of the MRS
of the US Senate Google correlation network during the election
period. A radical structural rebuilding of the MRS was observed during
this period and was actually quite surprising because the webpages searched
using Google are not always about the current news topics, but more like
archives of the WWW from the entire historical database. The radical movements of senators in the MRS show that
the dynamical webpages such as news articles, blog entries, and Wiki
pages take a considerable amount of space on the
WWW~\cite{BernersLee2006}.

The most outstanding rearrangement in this period is a great
movement of Senators from John McCain's side to John Kerry's side
on November (Figs.~\ref{ElectionUnited}(b) and (c)). The
movement of the Republican election candidates (whether the candidate
was re-elected or not) is particularly interesting. We suspect that one of the main
reasons for this major change of the MRS is Senator John Kerry's
``botched joke''
about the Iraq War on October 30
and the following
controversy~\cite{KerryBotchedJoke}. Besides the MRS from Google correlation values
we used, the impact of John Kerry's joke
can also be checked in Google Trends, with which one can find how
often people have searched certain topics on Google over
time~\cite{GoogleTrends}. We believe that many Republicans, who were at
John McCain's side in the MRS before the elections
(Fig.~\ref{ElectionUnited}(a)) were involved in the controversy
(with election candidates being most active), and their maximum
Google correlation values moved from that with John McCain to that
with John Kerry. After the elections, the impact of the controversy
was relatively weakened and the MRS was reshaped again
(Fig.~\ref{ElectionUnited}(d)). Although we have only discussed the
major movement tendency of senators and one possible cause, many
other interpretations and further studies are possible.
The techniques of Google correlation and MRS are widely applicable, and
further progress will be achieved in the future.

\subsection* {Aids to Obtain Further Specific Information}
Relatedness, quantified by the Google correlation, could be the
concept from either cooperation or competition. Google correlation
values cannot solely distinguish whether a given relationship is
friendly or hostile. External information can help us to
specify the relationships in more detail, and, this this section, we show an example of
such a specification with the US Senate network. The
record of Roll Call Votes of the US Congress (http://thomas.loc.gov/home/rollcallvotes.html),
which
guarantees that every senator's vote is recorded, is used to
elaborate relationships among senators.

With 642 Roll Call Votes of senators in the 109th
Congress, we assign the {\em vote correlation}
value $C(i,j)$ for every pair of senators $i$ and $j$ as follows:
\begin{equation}
C(i,j) = \frac{\displaystyle \sum_{n} X_n (i,j)}
{\displaystyle \sum_{n} \delta_n (i) \delta_n
(j)},\label{VoteCorrelation}
\end{equation}
where $X_n (i,j)$ is $1$ if senator $i$ and $j$ concurrently voted
for or against the bill of the $n$th Roll Call Vote and $-1$
otherwise, and $\delta_n (i)$ is $1$ if Senator $i$ participated in
the $n$th Roll Call Votes and $0$ if Senator $i$ did not
vote. We exclude the cases of unanimous votes to
remove the effect of the entire Senate's opinion. Then, $C(i,j) \in [-1, 1]$ and measures
the correlation of opinions of senator $i$ and $j$.

Now we can infer the degree of cooperation with the vote correlation
defined in Eq.~(\ref{VoteCorrelation}).
In Fig.~\ref{Senator_MRS},
we distinguish the links among senators with the positive and
negative vote correlation. From Fig.~\ref{Senator_MRS}, we
observe that the positive vote correlation is almost always given to
the senator pairs from the same party and the negative vote
correlation to the senator pairs from the different parties. Among
all the senator pairs, only $5.66\%$ are from the different
parties and have positive vote correlation value and $0.08\%$ are with
the same party and have negative vote correlation value, which implies
the partisan polarization discussed in Ref.~\cite{YZhang2007}.

\subsection*{Relationships between Two Groups: Bipartite Network Analysis}

Investigating relationships via search engines is not restricted to
a specific group of people. In addition, objects in a search query do
not have to be restricted to people's names. We demonstrate this
fact by investigating the relatedness between politicians and large
corporations, revealing possible connections between politics and
business. For sets of politicians, we selected 18 potential US
presidential candidates in January
2008~\cite{CandidateList}
and the 109th US Senators. We
chose the 100 largest corporations, as reported by
{\em Fortune}~\cite{Fortune}
as the set of corporations.

The method of analysis is similar to the previous one, but in this
case Google correlation values only {\em between politicians and
corporations} are considered in a way to construct a so-called
``bipartite network.'' MRS is generated by collecting links from
politicians to the corporations to which they are related most and
vice-versa. Another measure introduced is the
normalized Google correlation~\cite{NormalizedGoogleCorrelation}
which represents the relatedness where the effect of fame is
removed. This new measure is able to effectively prevent
famous nodes from ``dictating'' the network.
All the data for this analysis were collected in January 2008.

\begin{figure}[!ht]
\begin{center}
\includegraphics[width=0.9\textwidth]{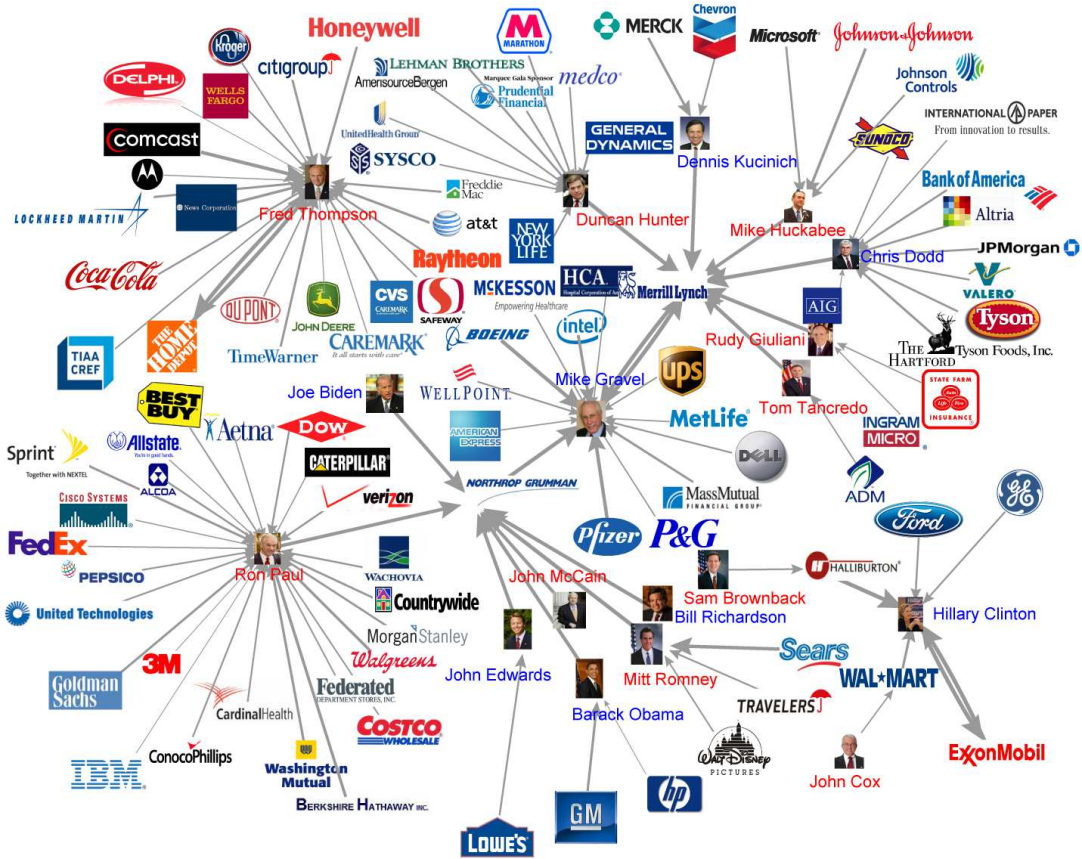}
\end{center}
\caption{
{\bf MRS from the bipartite network of the US presidential
candidates and the 100 corporations.} The Democratic candidates'
names are colored as blue, the Republican candidates' names as Red, and the
corporations as their logos. Normalized Google correlation values
are used.
}
\label{CandCorp_NOR_MRS}
\end{figure}

Figure~\ref{CandCorp_NOR_MRS} shows the MRS from the normalized
Google correlation network of the US presidential candidates and the
100 corporations. John McCain, who has become the actual Republican
presidential candidate at the time of writing, does not have many
connections with large corporations in MRS. However, the only connected
corporation with him is Northrop Grumman, which recently won the
joint tanker contract to assemble the KC-45 refueling tankers for the US
Air Force with EADS~\cite{EADS}. Because Senator John
McCain once uncovered a corrupt effort by Boeing, which is Northrop
Grumman's rival company~\cite{NorthropGrumman}, the connection looks
interesting. The thick bidirectional connection between Senator
Hillary Clinton and Exxon Mobil is likely from the large amount of
money contributed to Senator Clinton from the
corporation~\cite{ExxonMobil}. In similar ways, such analysis might
give some hints for further investigation for the relationship
between politics and business.

\begin{figure}[!ht]
\begin{center}
\includegraphics[width=0.9\textwidth]{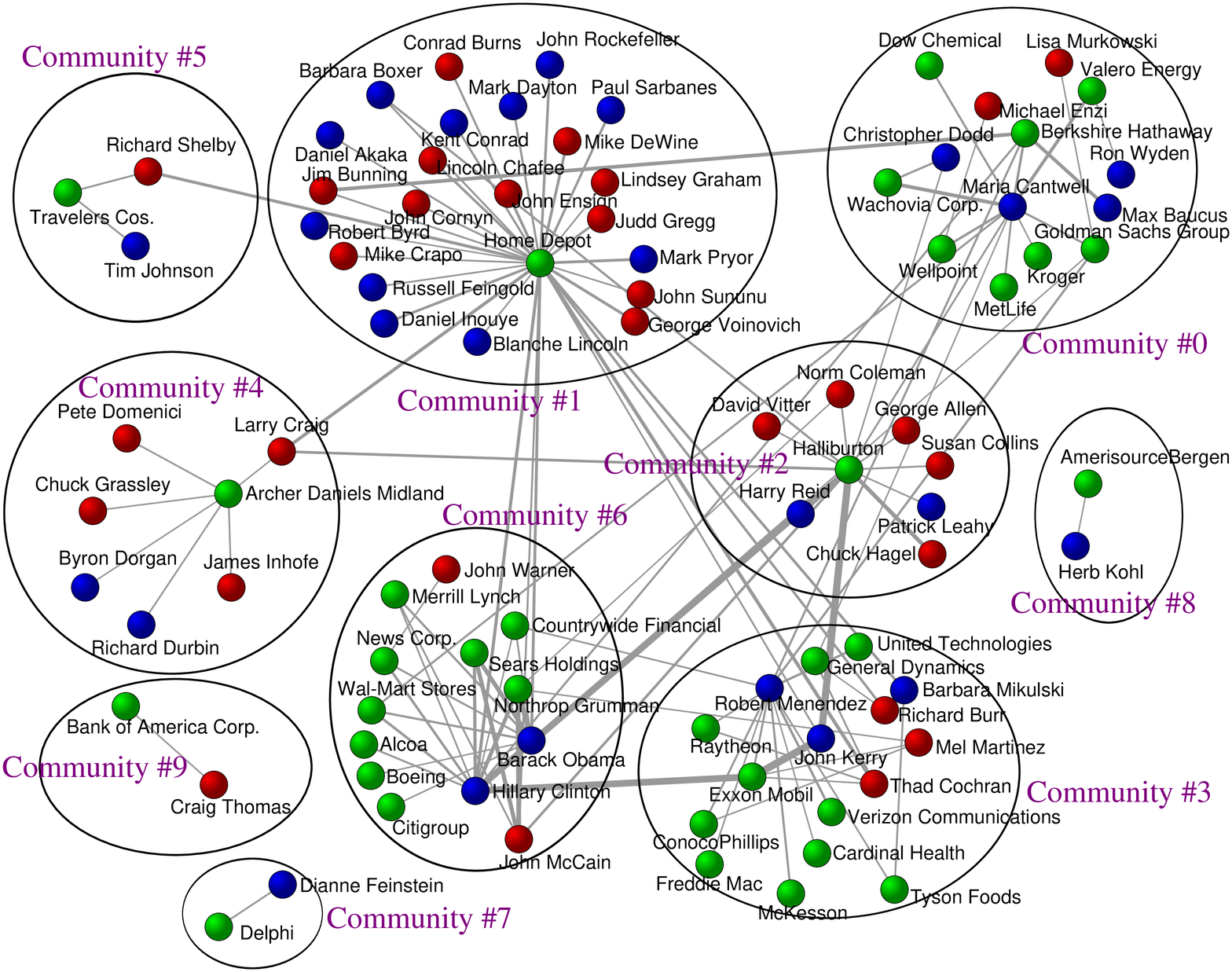}
\end{center}
\caption{
{\bf Community structure of the subnetwork.} We keep only
normalized Google correlation values larger than $0.002$,
from the normalized Google correlation network of the 109th US
Senators and the 100 corporations. The Democratic Senators are
colored as blue, the Republican Senators as Red, and the
corporations as green.
}
\label{NOR_CorpSen_2m_comm}
\end{figure}

We also tried to elucidate community structures from the bipartite network
between politicians and corporations as shown in Fig.~\ref{NOR_CorpSen_2m_comm}.
First we extracted the
normalized Google correlation values between US Senators and the 100
corporations. Then we kept the link, whose Google correlation value
is larger than $0.002$, to obtain a sparser subnetwork for visualization. The community
structure from the subnetwork was obtained by Newman's eigenvalue
spectral method~\cite{Newman2006} and the modularity $Q$ is $0.58$,
which might reveal the subunit of politics-business connections.

\subsection*{Comparison with Real Social Networks}
In this section, we provide evidence for the validity of
social network construction by Google correlation values. We
obtained a scientific collaboration network~\cite{WebOfScience}
among the authors of papers citing the five key
papers~\cite{Newman2003a,Albert2002,Dorogovtsev2001a,Watts1998,Barabasi1999}
in the network theory. The 776 authors who wrote at least three
papers were selected due to computational tractability. In this
collaboration network, the pairs of authors who wrote the papers
together were connected and the weights were assigned as the numbers
of collaborated papers. To test the reliability of the Google
correlation network among these authors, we constructed a
weighted social network with the Google correlation
values~\cite{AddedNetwork}.

\begin{figure}[!ht]
\begin{center}
\includegraphics[width=0.5\textwidth]{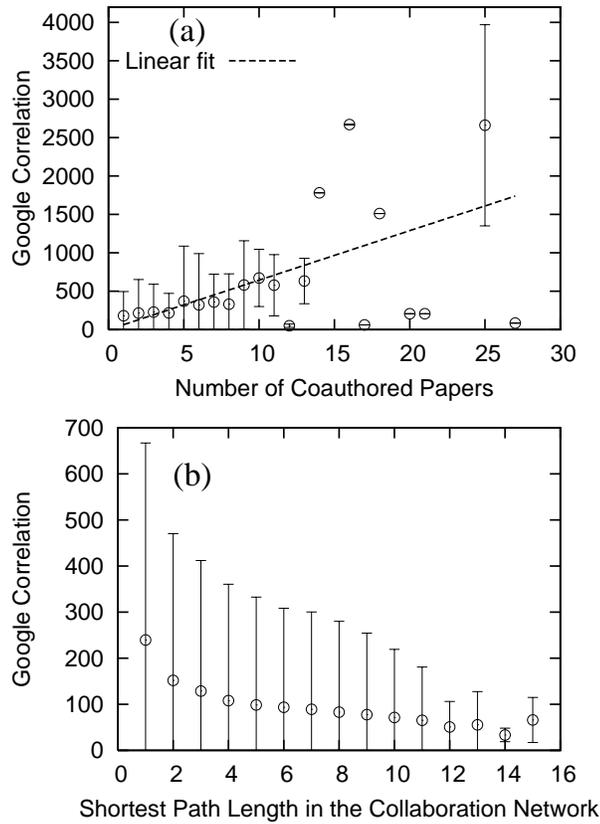}
\end{center}
\caption{
{\bf Comparison of a real social network and the Google correlation network.}
(a) The average Google correlation values for each number of
collaborated papers. The error bars represent the standard
deviations. The Pearson correlation coefficient between the number
of collaborated papers and Google correlation values is $0.268$ and
the dashed line is the linear fit whose slope is $64.4$.
(b) The average Google correlation values for each value of the
shortest path length among pairs of nodes in the collaboration
network. The error bars represent the standard deviations. The
Pearson correlation coefficient between the shortest path length and
the Google correlation values is $-0.115$.
}
\label{CoauthorGoogle}
\end{figure}

The direct comparison between these two weighted networks (the
collaboration network and the Google correlation network) is
nontrivial, partly because of the enormous difference in the link
density, i.e., the collaboration network is much sparser. Therefore,
we suggest two schemes for comparison. First, we check the
correlation between the weight in the collaboration network (the
number of collaborated papers) and the Google correlation values for
pairs of connected authors in the collaboration network. If the
Google correlation network represents the true relatedness, we
expect a positive correlation between the two quantities and
Fig.~\ref{CoauthorGoogle}(a) indeed shows a positive correlation.
Second, regardless of whether two nodes in the collaboration network are
directly connected or not, the Google correlation value and the
shortest path length in the collaboration network for those two
nodes are expected to be negatively correlated.
Figure~\ref{CoauthorGoogle}(b) confirms this expectation. Because the Google
correlation value represents the relatedness of two authors, the
larger the Google correlation value of the two authors, the nearer
they are located in the collaboration network.

These correlations, of course, are not perfect. However, we suggest
that the difference does not indicate the error or limitation of the
Google correlation but reveals the actual difference between the
collaboration and relatedness. Two authors can have large Google
correlation value, even if they have never written papers
together, if they work in the similar fields, show up at the same
conferences many times, and thereby appear in the same ``participant
list'' webpages of many conferences, for example. In summary, we
have verified that our method actually reflects the structure of the
real coauthorship network and have demonstrated the potential of our
method.

Finally, we should mention caveats of our method. Many webpages are not under the quality control and may contain misleading or alleged facts. Therefore, our method should be considered as a {\em proxy} reflecting the real correlations. In other words, one has to be careful when dealing with the Google correlation data and note that any conclusions drawn from the analysis should be followed by accurate follow-up investigations, like genome-wide computational predictions followed by high-quality, small-scale experiments in biology.

However, in any case, we would like to emphasize that the Google correlation values can be the first, useful and exploratory step towards further investigations. We also want to point out that it is possible to flexibly customize the definition of the correlation measure for different purposes, for instance, by dividing the raw cooccurrence value by their Google hit values to get rid of their popularity effects whenever it is necessary, as suggested in the previous sections. Another way to customize our method is to use more specific search engines. For instance, for the coauthorship relations, one can count cooccurrences from Google Scholar, which indexes only the scholarly literature. Public relationships among politicians can be extracted more accurately by focusing on only the news articles. As an example, we constructed a network of Korean politicians by counting the number of news articles from a Korean online news service~\cite{Naver}, and demonstrated that the two clear groups in MRS well correspond to political parties and each party's leader/influential person possessing central position with many incoming links.

\section*{Discussion}
There is a tremendous amount of data on the Web, which can prove very
useful if we harness it cleverly. Search engines are a
basic device to classify such information and we have constructed
social networks based on the Google correlation values quantifying
the relatedness of people. We have systematically analyzed the basic statistical
properties from the viewpoint of weighted network theory, introduced
a new quantity called the R{\'e}nyi disparity to represent the
different aspect of the weight distribution for individual nodes,
and suggested MRS to elucidate the essential relatedness. We have
used the US Senate as a concrete example of our analysis and presented
the results.

The concepts of the R{\'e}nyi disparity and MRS introduced in this
paper are not restricted to the Google correlation network, of
course. The process of finding out ``hidden asymmetry'' of weighted
links is applicable to other many weighted networks from various
disciplines as well. In other words, such concepts can be
interpreted as useful characteristics in different contexts. We have also
compared a real scientific collaboration network with the
social network constructed by our method introduced in this paper
and discussed the result. The larger Google
correlation values two authors have, the more papers they tend to
have written together, causing them to appear to be ``closer''
in the scientific collaboration network.

Extracting information on the Web to construct networks makes it
possible not only to obtain large networks with many participants,
but also to monitor the change of such networks by
collecting data on a regular basis. We have verified that the network
structures do not change abruptly, partly because the Web plays the
role of a digital ``archive,'' not a ``newspaper.'' However, during
important events such as the elections for the United States
Senate held in November 2006, the US Senate network was
significantly reformed as we have discussed in this paper. If the
webpages were classified into several categories such as news
articles, blog articles, {\em etc.}, more information would be available.
We hope that so-called Web 2.0~\cite{Henzinger2007,BernersLee2006,NatureSpecialIssue,Henzinger2004}
will significantly increase the possibility to obtain such
classified information with ease in the future. The proper use of
the Web and search engine in scientific research has already
begun, for instance, in the research on the human tissue-specific
metabolism~\cite{Shlomi2008}, and we welcome other researchers
who will join this movement in the future.


\section*{Acknowledgments}
We thank Daniel Kim for building the data extraction platform
with the Google Search API.
This work was supported by NAP of Korea Research Council of
Fundamental Science \& Technology.


\begin{thebibliography}{99}
\bibitem{AddedNetwork}
To avoid the ambiguity of authors' name,
the word ``network'' is added to the search query in this case,
assuming most authors are related to the network research.
\end{thebibliography}



\end{document}